\documentclass[aps,prb,twocolumn,showpacs,superscriptaddress]{revtex4-1}
\usepackage{graphicx}

\begin{document}

\title{Strain-activated structural anisotropy in BaFe$_2$As$_2$}

\author{Xiang Chen}
\affiliation{Department of Physics, Boston College, Chestnut Hill, Massachusetts 02467, USA}
\affiliation{Materials Department, University of California, Santa Barbara, California 93106, USA.}
\author{Leland Harriger}
\affiliation{NIST Center for Neutron Research, National Institute of Standards and Technology, Gaithersburg, MD 20899, USA}
\author{Athena Sefat}
\affiliation{Materials Science and Technology Division, Oak Ridge National Laboratory, Oak Ridge, Tennessee 37831, USA}
\author{R. J. Birgeneau}
\affiliation{ Physics Department, University of California, Berkeley, California 94720, USA}
\affiliation{ Materials Science Division, Lawrence Berkeley National Lab, Berkeley, California 94720, USA }
\affiliation{ Materials Science Department, University of California, Berkeley, California 94720, USA}
\author{Stephen D. Wilson}
\email{stephendwilson@engineering.ucsb.edu}
\affiliation{Materials Department, University of California, Santa Barbara, California 93106, USA.}

\begin{abstract}
High-resolution single crystal neutron diffraction measurements are presented probing the magnetostructural response to uniaxial pressure in the iron pnictide parent system BaFe$_2$As$_2$.  Scattering data reveal a strain-activated, anisotropic broadening of nuclear Bragg reflections, which increases upon cooling below the resolvable onset of global orthorhombicity.  This anisotropy in lattice coherence continues to build until a lower temperature scale---the first-order onset of antiferromagnetism---is reached.  Our data suggest that antiferromagnetism and strong magnetoelastic coupling drive the strain-activated lattice response in this material and that the development of anisotropic lattice correlation lengths under strain is a possible origin for the high temperature transport anisotropy in this compound.
\end{abstract}


\maketitle
\section{Introduction}
Studies of the electronic properties of iron-based high temperature (high-$T_C$) superconductors have identified an unusually large susceptibility to uniaxial strain associated with the low temperature tetragonal to orthorhombic structural distortions ($T_S$) present in their parent and doped variants \cite{0034-4885-74-12-124506}.  This distortion is commonly associated as a secondary manifestation of electronic order, driven by either spin or orbital degrees of freedom \cite{FernandesReview, PhysRevB.85.024534}.  Above this transition, anomalous electronic anisotropies are observed under the application of symmetry breaking fields in the nominally high-temperature tetragonal state \cite{Chu10082012, Yi26042011, 0034-4885-74-12-124506, Kasahara}.  Specifically, a large susceptibility of the tetragonal state to  $C_4$ rotational symmetry breaking under biasing fields has inspired the proposal of an otherwise hidden electronic nematic state in many of these systems \cite{0034-4885-74-12-124506}---a phase with potential relevance to the formation of their high-$T_C$ ground states \cite{kivelson}.     

Electronic order and modifications to the underlying crystallographic structure are necessarily intertwined in iron pnictide superconductors, where the symmetry of the tetragonal lattice must be lowered in order to accommodate the formation of stripe-like, long-range antiferromagnetism (AF) \cite{delacruz, PhysRevLett.101.257003}. Measurements have unveiled that, at temperatures above the global orthorhombic distortion at $T_S$, an apparent electronic nematicity appears in single domain samples \cite{Chu13082010}.  One mechanism for this effect is a shifting of the onset of the structural distortion upward in temperature under the biasing fields necessary for the preparation of a single domain sample \cite{PhysRevLett.108.087001,PhysRevB.85.144509}.  In particular, the use of uniaxial strain to remove twin domains \cite{PhysRevB.79.180508} and probe in-plane anisotropies simultaneously promotes both orthorhombicity and magnetic order at temperatures conventionally envisioned in the paramagnetic, tetragonal regime \cite{PhysRevB.85.144509, PhysRevLett.108.087001, PhysRevB.89.214404}.  The primary manifestation of this is the strain field's breaking of the in-plane C$_4$ symmetry and smearing of $T_S$ into a gradual crossover whose distortion builds upon cooling \cite{PhysRevLett.108.087001, PhysRevLett.115.197002}.  

Notably, the microscopic means through which this strain-driven crossover builds and its mechanism for driving the observation of electronic nematicity remain open questions.  For instance, while uniaxial strain is known to renormalize magnetostructural phase behavior such as decoupling the antiferromagnetic and structural phase transitions \cite{PhysRevB.87.174503, PhysRevB.89.214404}, the relationships between strain-induced orthorhombicity, correlated magnetism, and electronic anisotropy remain difficult to access.  The bilayer iron pnictide parent system BaFe$_2$As$_2$ (Ba-122) is a relatively simple platform with a prominent strain-activated nematic response \cite{PhysRevB.81.184508} well suited for exploring this question; one where compressive pressure applied along the orthorhombic $b$-axis has already shown dramatic effects in the magnetostructural phase behavior \cite{PhysRevLett.108.087001}.   

In this paper, we examine Ba-122's lattice response to strain via a high-resolution neutron diffraction study.  Consistent with earlier reports, but contrary to a recent study reporting no effect\cite{Lu08082014}, compressive strain applied along the short, in-plane $b$-axis dramatically shifts the resolvable orthorhombicity to higher temperatures promoting a nematic response.  Strikingly, our data reveal an anisotropy in the structural coherence or crystallinity of the sample that develops only under uniaxial strain in the proposed nematic regime.  This appears in the form of an anisotropy in the structural Bragg widths that builds as the magnetic ordering temperature is approached.  We propose a picture where the anisotropy in lattice correlation lengths arises via strain-biased lattice twinning at high temperatures that is eventually relaxed by the onset of long-range magnetic order.  Our results demonstrate the importance of magnetoelastic coupling and local lattice symmetry breaking in driving the nematic response of BaFe$_2$As$_2$.

\begin{figure}
\includegraphics[scale=.25]{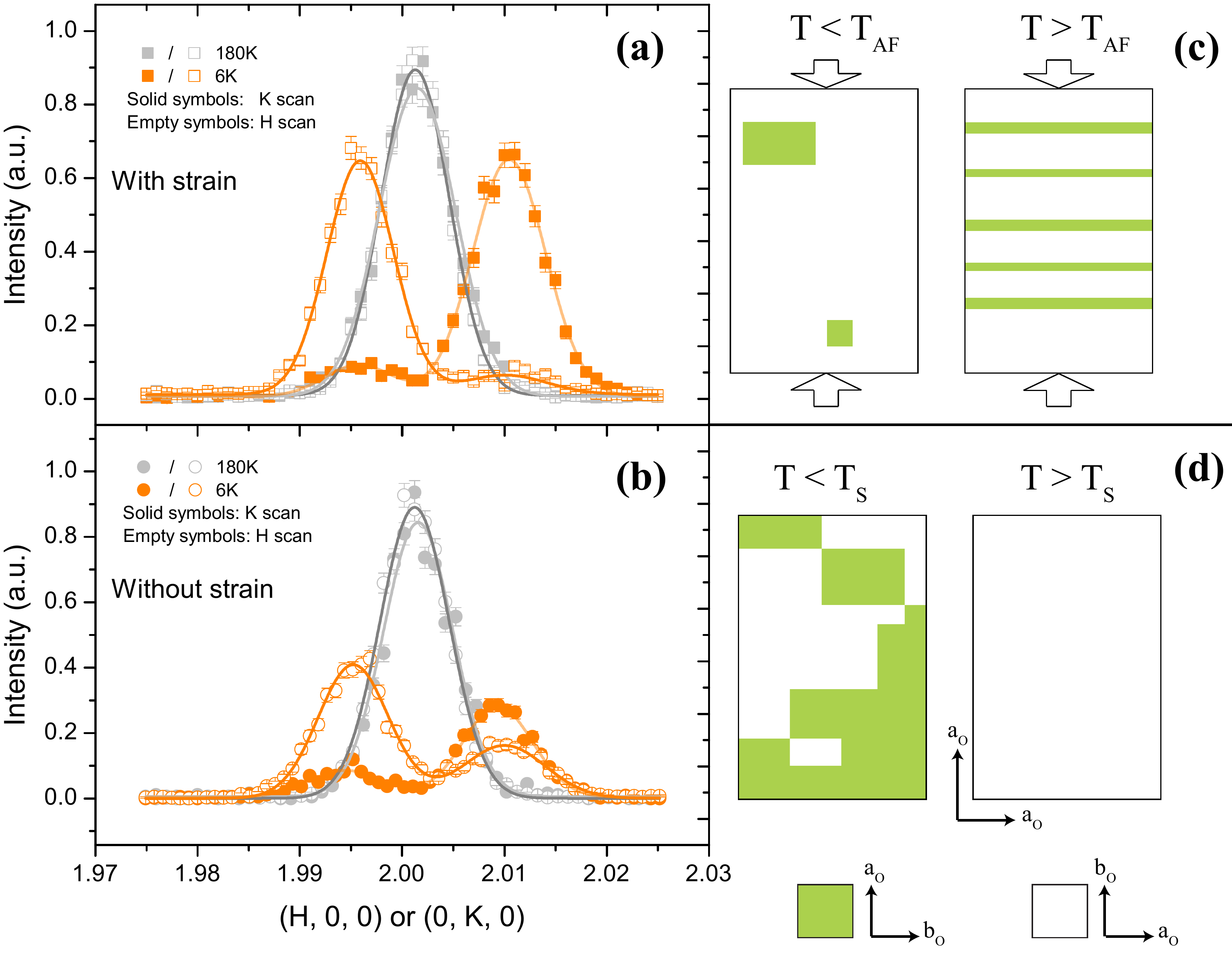}
\caption{(a) Radial scans through the \textbf Q=(2, 0, 0) (open squares) and \textbf Q=(0, 2, 0) (closed squares) positions under applied uniaxial pressure along the $b$-axis.  Data is shown at $180$ K (grey symbols) and $6$ K (orange symbols). (b) Radial scans through the \textbf Q=(2, 0, 0) (open squares) and \textbf Q=(0, 2, 0) (closed squares) positions under zero strain.  Data is shown at $180$ K (grey symbols) and $6$ K (orange symbols). (c) and (d) illustrate the domain distribution and anisotropies suggested by the aggregate structural data.  Green and white domains illustrate a simple picture of anisotropy in twin densities under different strain states and at different temperatures as seen projected into the $ab$-plane.  The compressive strain direction is denoted by the arrows which define the shorter orthorhombic b-axis}
\end{figure}

\begin{figure}
\includegraphics[scale=.34]{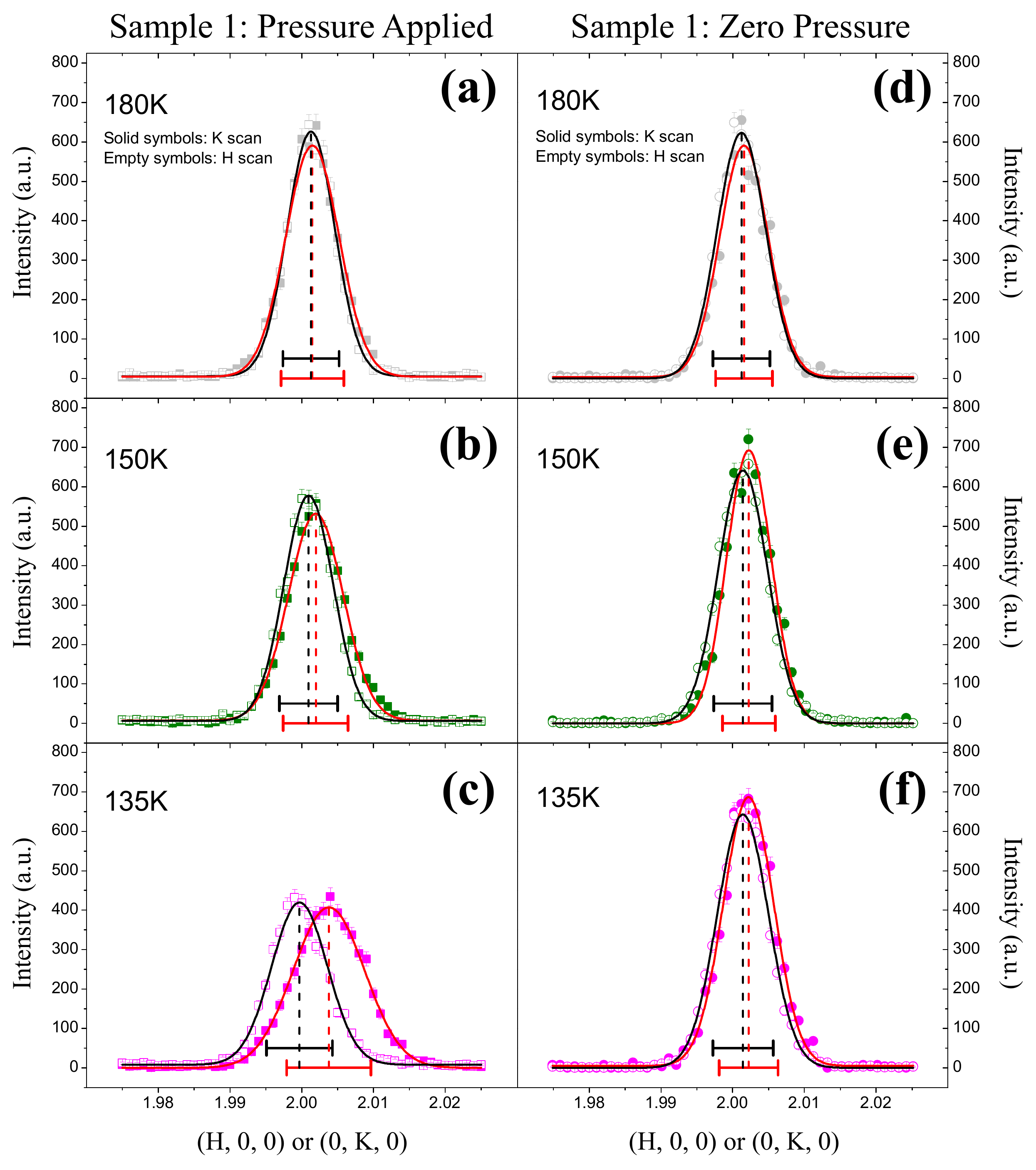}
\caption{ Radial $H$- and $K$-scans through the (2,0,0) and (0,2,0) positions under applied uniaxial pressure (a-c) and after releasing pressure (d-f).  Open symbols denote scans through the (2,0,0) and closed symbols are through the (0,2,0).  Data is shown for scans collected at 180 K (a,d), 150 K (b,e), and 135 K (c,f).  Bars centered beneath peaks denote the FWHM of the Gaussian fits as described in the text.  Vertical dashed lines denote fit peak centers.}
\end{figure}

\section{Experimental Details}
Neutron scattering is an important tool for exploring the lattice response to uniaxial pressure due to its ability to sample the entirety of the sample volume and resolve anisotropies averaged across the sample.  It provides a window where the bulk response to the mean strain field can be isolated and inhomogenous strain fields are effectively averaged out.  Our neutron diffraction data was collected on the SPINS spectrometer at the NIST Center for Neutron Research.  A single crystal of Ba-122 (dimensions $\approx 1.5 \times 1.0 \times 0.2$) mm was cut with facets parallel to the orthorhombic $a$ and $b$-axes and loaded into a pressure clamp identical to earlier studies \cite{PhysRevLett.108.087001}.  Compressive uniaxial pressure was applied parallel to the $b$-axis at room temperature, the clamp was then mounted in a helium closed cycle refrigerator, and the sample aligned within the [H, K, 0] scattering plane. Due to its enhanced susceptibility to uniaxial strain, an as-grown crystal of Ba-122 with a slightly suppressed $T_S\approx 132$ K was chosen for our experiments.  Post-growth annealing is known to enhance $T_S$ to 140 K and to harden the lattice \cite{PhysRevB.82.144525, PhysRevB.87.184511}---largely masking anisotropies driven under modest strain fields \cite{Liang2011418}. Fixed neutron energies of $E_i=E_F=5$ meV were selected using the (0,0,2) reflections of pyrolitic graphite (PG) monochromator and analyzer crystals.  No filters were utilized, so scattering data also contains higher harmonics of the incident beam.  For our purposes, this means that reflections indexed as (2, 0, 0)/(0, 2, 0) are dominated by the much stronger (4, 0, 0)/(0, 4, 0) Bragg scattering via $4E_i$.  Scattering wave vectors in this paper are indexed using the lattice using the larger, orthorhombic unit cell of Ba-122.  

\section{Neutron diffraction results}
Looking first at Fig. 1, the effect of an in-plane strain field in partially detwinning a Ba-122 crystal ($P1\approx 0.7$ MPa) can be readily observed.   Figs. 1 (a) and (b) show low temperature $6$ K radial scans through the $(2, 0, 0)$ and $(0, 2, 0)$ positions both in zero pressure and under applied pressure.  Below the zero strain distortion temperature, $T_S$, a clear splitting in each reflection is observed due to macroscopic twinning of the crystal as the tetragonal symmetry is lifted.  The inequivalent population of the twin domains under zero pressure is likely due to a small inherent strain field imparted during sample mounting or frozen in during the sample's growth.  However, once a $b$-axis oriented pressure is applied, the low temperature twinning is largely suppressed and the sample is nearly detwinned as shown in Fig. 1 (a).  The twinning ratio---the volume occupying one minority twin/set of twins over the majority twin volume---at base temperature changes from $\approx 33\%$ in the nominally strain-free case to $\approx 10\%$ in the strained state.

Under this modest pressure (i.e. below the threshold for complete detwinning), the magnitude of the saturated orthorhombic distortion at 6 K is increased by $\approx 12.6\pm 0.1 \%$, and the transition is broadened into a crossover where the orthorhombicity of the lattice is resolved below $240$ K.  For small distortions, the absolute magnitude of orthorhombicity $\delta=\left| \frac{a - b}{a+b} \right|$ lies within the resolution of the instrument; however the onset of orthorhombicity can still be isolated as the temperature at which the thermal gradients of the unique, strain-defined, $a$ and $b$ lattice constants become distinguishable.  Small differences in the in-plane lattice constants were observed even at room temperature; however these offsets are within the absolute resolution of the spectrometer and are observed in the sample both under zero and applied pressure.  Under strain this offset was $a$(300 K)=$5.6064$ $\AA$, $b$(300 K)=$5.6078$ $\AA$, and with pressure released this offset was $a$(300 K)=$5.6015$ $\AA$, $b$(300 K)=$5.6009$ $\AA$.  For clarity, we have removed this high temperature offset and normalized $a$- and $b$-lattice constants to be equivalent at room temperature; however our measurements do not preclude a small strain-induced room temperature orthorhombicity such as that reported in a recent Larmor diffraction study \cite{2015arXiv150704191L}.    

Momentum scans through the $(2, 0, 0)$ and $(0, 2, 0)$ positions collected above the strain-free $T_S=132$ K are plotted in Fig. 2 for both cases of applied and released uniaxial pressure.  Once uniaxial pressure is applied, overlaid peak centers for $H$- and $K$-scans shift in opposite directions as the sample is cooled, and peaks in this high temperature regime remain well characterized by single Gaussian line shapes down to the zero strain $T_S$.  After pressure is released, the sample relaxes back to the phase behavior expected for the weakly first order magnetostructural phase transition, and peak positions and line shapes remain unchanged in this high temperature regime.  

Fig. 3 shows the evolution of Bragg peak line shapes and positions fit to the form $A e^\frac{-(x-x_0)^2}{2w^2}$ as the sample is cooled both under applied pressure and with the pressure released.  Once pressure is applied (Fig. 3 (a)), the orthorhombic distortion is resolvable around $\approx 240$ K.  For comparison under zero pressure, Fig. 3 (b) shows the onset of the orthorhombic distortion at $132$ K.   In this zero pressure case, the apparent full width at half maxima (FWHM$=2w\sqrt{2ln(2)}$) of the Bragg peaks at the $(2, 0, 0)$ and $(0, 2, 0)$ positions are plotted in Fig 3 (d).  They uniformly diverge, sharply, at $T_S$ as the lattice distorts and twins are convolved within one Gaussian peak.  In contrast however, Fig. 3 (c) reveals that once strain is applied the widths of these Bragg peaks become increasingly asymmetric upon cooling.  

\begin{figure}
\includegraphics[scale=.375]{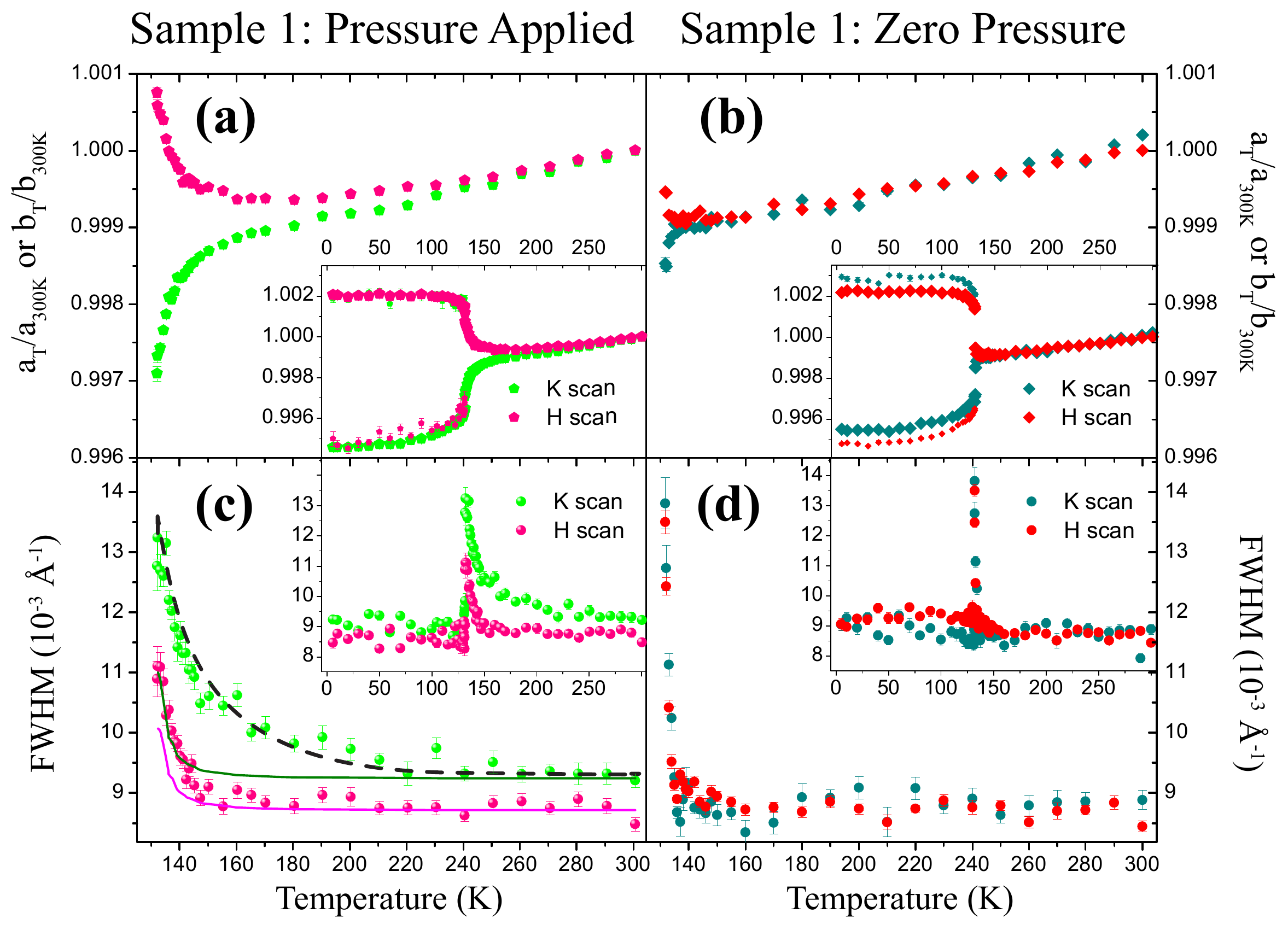}
\caption{Evolution of the (2,0,0) and (0,2,0) structural peaks as a function of temperature both with and without externally applied strain.  Panels (a) and (b) show the evolution of lattice parameters as the sample is cooled under strain and with strain released respectively.  Insets show an expanded range of the same data. Panels (c) and (d) plot the corresponding peak widths in both strained and strain-free states.  Solid lines show the fits to a purely twinning driven model of peak broadening as described in the text, and insets again plot an expanded range of the data.  Dashed line is a guide to the eye showing the divergence of structural anisotropy approaching $T_S$.}
\end{figure}

Looking closer at data collected under strain in Fig. 3 (c), an initial anisotropy in the widths of the $(2, 0, 0)$/$(0, 2, 0)$ reflections is observed at room temperature.  This small asymmetry in peak widths at $300$ K likely arises due to trivial strain-broadening originating from the applied pressure, and the magnitude of this effect is comparable at both high ($300$ K) and low ($4$ K) temperatures.  As the system is cooled between these extremes however, the asymmetry in the $H$- and $K$- axis peak widths surprisingly grows upon cooling.  Specifically, the inherent width of $(0, 2, 0)$ peak broadens relative to that of the $(2, 0, 0)$ until a lower energy scale at $132$ K is reached. Below this temperature a clear twinning is resolved and the inherent Bragg widths of both orthorhombic domains relax to their high temperature values.  

Above $132$ K, the anisotropy in Bragg widths can be discriminated from a trivial twinning-driven effect as the orthorhombicity increases (ie. a twin convolving with the majority peak's profile and creating the appearance of broadening).  Naively, a twinning-driven broadening should be symmetric and, furthermore, it can be modeled using the known twinning ratio and temperature dependent lattice constants under strain. The expectations of the thermal evolution of the $(2, 0, 0)$ and $(0, 2, 0)$ peak widths using this model are plotted as solid lines in Fig. 3 (c).  Here trivial twin formation can account for the apparent increase in the Bragg width of the $(0, 2, 0)$ reflection upon cooling; however it fails to account for the increasing anisotropy in the $H$ (i.e. $(2, 0, 0)$) and $K$ (i.e. $(0, 2, 0)$) structural correlation lengths observed upon cooling between $240$ K$\geq T\geq 132$ K.  

\section{Discussion}
The persistence of a sharp lattice response below $132$ K in the presence of a biasing strain field is unusual.  For instance, our data show that by $150$ K a global lattice orthorhombicity is established, yet the anisotropy in the Bragg widths continues to build upon continued cooling toward $132$ K before promptly relaxing. Previous experiments have shown that neutron scattering intensity changes typically associated with extinction release upon cooling through $T_S$ are also largely unaffected by the application of uniaxial strain \cite{Lu08082014}.  This occurs despite the shift of significant orthorhombicity to much higher temperatures, and it supports the notion of lattice coupling to a remnant energy scale which remains more sharply defined under strain.    

AF order in Ba-122 couples to strain in a secondary manner \cite{PhysRevB.89.214404} and is the likely source of the remnant energy scale driving the lattice at $132$ K \cite{Supplemental}.  The first-order nature of the AF transition allows it to largely survive sharply defined under the application of strain.  One scenario where AF order can drive the observed lattice anisotropy is via magnetoelastic coupling between short-range magnetic correlations and twin formation above the nominal $T_S$---the latter of which can be readily biased via the application of an external strain field \cite{PhysRevLett.108.087001, PhysRevB.89.214404}.  For instance, structural twin density may progressively increase along the short $b$-axis upon cooling in order to shield the applied strain field as the system approaches the lattice instability at $T_S$ (Figs. 1 (c) and (d)) \cite{PhysRevLett.105.157003}.  As it does so, there is an increasing surface energy cost, a portion of which is due to moments frustrated along twin boundaries. This balance can naively be biased as long-range magnetic order is established and magnetoelastic coupling drives a restructuring of the twin domain volumes into mesoscopic twins.  Local structure measurements have previously reported nanotwins above $T_S$ in Ba-122 \cite{PhysRevB.85.224506, PhysRevB.88.184515, PhysRevB.86.174113}, further suggesting twin dynamics under strain may play a role in driving the lattice anisotropy.    

To further explore the role of strain in generating the observed lattice anisotropy, a second sample from the same growth batch was explored with increasing levels of applied uniaxial pressure. Data from this second sample are plotted in Fig. 4 where two pressures, $P1$ $(\approx 0.5$ MPa) $< P2$ $(\approx 0.7 $MPa) were applied.  While the smaller applied pressure $P1$ biased a resolvable onset of orthorhombicity up to $170$ K (reflecting a lower resulting strain field than that applied in Fig. 3), no high temperature anisotropy in Bragg correlation lengths was resolved.  Upon increasing the pressure to $P2$, the apparent onset of orthorhombicity continues to shift upward to $\approx 250$ K and a high temperature anisotropy in Bragg correlation lengths emerges.  This correlates the magnitude of the strain field within the sample to the degree of anisotropy and demonstrates that the effect is not simply inherent to fluctuations accessed via an externally broken $C_4$ symmetry.  Further supporting this claim, recent Larmor diffraction measurements have shown that the strain-induced shift in the onset of orthorhombicity is truly \textit{static} within $\approx 1$ $\mu$eV \cite{2015arXiv150704191L}.  

\begin{figure}
\includegraphics[scale=.375]{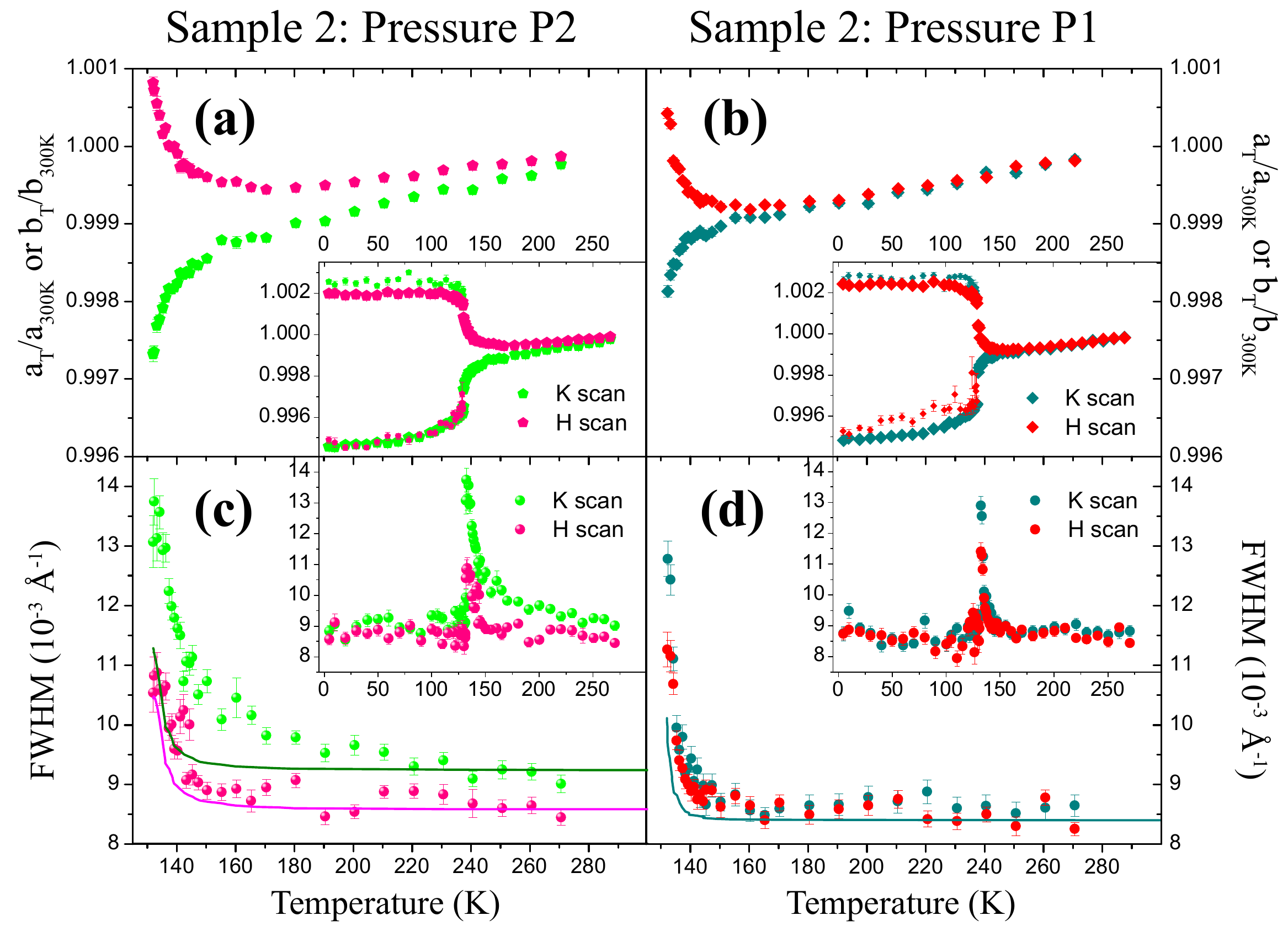}
\caption{Evolution of the (2,0,0) and (0,2,0) structural peaks as a function of temperature under increasing applied pressure.  Panels (a) and (b) show the evolution of the lattice parameters as the sample is cooled under applied pressure P1 and P2 respectively.  Insets show an expanded range of the same data. Panels (c) and (d) plot the corresponding FWHMs under P1 and P2 respectively.  Solid lines show the fits to a twinning driven model of peak broadening as in Fig. 3, and insets again plot an expanded range of the data.}
\end{figure}

Regardless of the detailed mechanism, our data demonstrate that the structural coherence or crystallinity in strained Ba-122 crystals is anisotropic at high temperatures.  At $150$ K, far from the nominal strain-free $T_S$, data in Fig. 3 (c) reveal a correlation length of $\xi \approx 479 \pm106\AA$ along the shorter, in-plane $b$-axis and resolution limited domains along the $a$-axis. This was calculated by deconvolving the experimental resolution (Gaussian width of $w_{res}= 0.0037 \AA^{-1}$) from the observed $(0, 2, 0)$ Bragg peak Gaussian width, $w$, and defining $\xi=w^{-1}\sqrt{2ln(2)}$. 

While the absolute orthorhombicity in this high temperature regime is small, this anisotropy in lattice correlation lengths can couple to electronic properties such as transport via increased scattering (i.e. higher resistance) along the compressive strain direction.   In other words, the anisotropy is reflective of a lattice whose periodicity is preferentially interrupted by defects/grain boundaries/domain walls in one direction over the other and will enhance charge scattering and increase resistivity along the direction of strain.  This effect will in principle occur in parallel with the known energy splitting of $d_{xz}$ and $d_{yz}$ orbitals, which under strain shifts to higher temperatures as $T_{S}$ is smeared \cite{Yi26042011}.  The magnitude of this splitting, however, remains static under the strain fields needed to detwinn Ba-122 crystals, whereas the transport anisotropy reported at $T_{AF}$ and above (into the nematic regime) continues to grow with increasing strain \cite{Man}.  This suggests that the lattice anisotropy and domain scattering effects at these temperatures play an important and potentially dominant role in driving the observed transport anisotropy.   

\section{Conclusions}
In summary, neutron diffraction data exploring the structural response to uniaxial pressure in Ba-122 have identified a strain-activated regime of anisotropic lattice correlation lengths apparent between the resolution of an orthorhombic lattice and the onset of AF order.  The magnitude of this anisotropy increases as the surviving AF transition is approached and vanishes below this energy scale, suggesting the formation of anisotropic domains whose densities evolve upon cooling via strong magnetoelastic coupling.  We envision the lattice effects reported here are more prominent in samples where higher strain fields needed to fully detwin crystals are introduced and that the resulting anisotropic lattice coherence or crystallinity suggests a mechanism for generating a nematic response.       

\acknowledgments{
This work was supported by NSF CAREER award DMR-1056625 (S.D.W.).  Partial support given by the US Department of Energy (DOE), Office of Basic Energy Sciences (BES), Materials Sciences and Engineering Division, (A.S.)  The work at Lawrence Berkeley National Laboratory was supported by the U.S. Department of Energy (DOE), Office of Basic Energy Sciences, Materials Science and Engineering Division, under Contract No. DE-AC02-05CH11231}

\bibliography{Ba122Bib}

\begin{thebibliography}{28}
\expandafter\ifx\csname natexlab\endcsname\relax\def\natexlab#1{#1}\fi
\expandafter\ifx\csname bibnamefont\endcsname\relax
  \def\bibnamefont#1{#1}\fi
\expandafter\ifx\csname bibfnamefont\endcsname\relax
  \def\bibfnamefont#1{#1}\fi
\expandafter\ifx\csname citenamefont\endcsname\relax
  \def\citenamefont#1{#1}\fi
\expandafter\ifx\csname url\endcsname\relax
  \def\url#1{\texttt{#1}}\fi
\expandafter\ifx\csname urlprefix\endcsname\relax\def\urlprefix{URL }\fi
\providecommand{\bibinfo}[2]{#2}
\providecommand{\eprint}[2][]{\url{#2}}

\bibitem[{\citenamefont{Fisher et~al.}(2011)\citenamefont{Fisher, Degiorgi, and
  Shen}}]{0034-4885-74-12-124506}
\bibinfo{author}{\bibfnamefont{I.~R.} \bibnamefont{Fisher}},
  \bibinfo{author}{\bibfnamefont{L.}~\bibnamefont{Degiorgi}}, \bibnamefont{and}
  \bibinfo{author}{\bibfnamefont{Z.~X.} \bibnamefont{Shen}},
  \bibinfo{journal}{Reports on Progress in Physics}
  \textbf{\bibinfo{volume}{74}}, \bibinfo{pages}{124506}
  (\bibinfo{year}{2011}).

\bibitem[{\citenamefont{Fernandes et~al.}(2014)\citenamefont{Fernandes,
  Chubukov, and Schmalian}}]{FernandesReview}
\bibinfo{author}{\bibfnamefont{R.~M.} \bibnamefont{Fernandes}},
  \bibinfo{author}{\bibfnamefont{A.~V.} \bibnamefont{Chubukov}},
  \bibnamefont{and}
  \bibinfo{author}{\bibfnamefont{J.}~\bibnamefont{Schmalian}},
  \bibinfo{journal}{Nat Phys} \textbf{\bibinfo{volume}{10}},
  \bibinfo{pages}{97} (\bibinfo{year}{2014}).

\bibitem[{\citenamefont{Fernandes et~al.}(2012)\citenamefont{Fernandes,
  Chubukov, Knolle, Eremin, and Schmalian}}]{PhysRevB.85.024534}
\bibinfo{author}{\bibfnamefont{R.~M.} \bibnamefont{Fernandes}},
  \bibinfo{author}{\bibfnamefont{A.~V.} \bibnamefont{Chubukov}},
  \bibinfo{author}{\bibfnamefont{J.}~\bibnamefont{Knolle}},
  \bibinfo{author}{\bibfnamefont{I.}~\bibnamefont{Eremin}}, \bibnamefont{and}
  \bibinfo{author}{\bibfnamefont{J.}~\bibnamefont{Schmalian}},
  \bibinfo{journal}{Phys. Rev. B} \textbf{\bibinfo{volume}{85}},
  \bibinfo{pages}{024534} (\bibinfo{year}{2012}).

\bibitem[{\citenamefont{Chu et~al.}(2012)\citenamefont{Chu, Kuo, Analytis, and
  Fisher}}]{Chu10082012}
\bibinfo{author}{\bibfnamefont{J.-H.} \bibnamefont{Chu}},
  \bibinfo{author}{\bibfnamefont{H.-H.} \bibnamefont{Kuo}},
  \bibinfo{author}{\bibfnamefont{J.~G.} \bibnamefont{Analytis}},
  \bibnamefont{and} \bibinfo{author}{\bibfnamefont{I.~R.}
  \bibnamefont{Fisher}}, \bibinfo{journal}{Science}
  \textbf{\bibinfo{volume}{337}}, \bibinfo{pages}{710} (\bibinfo{year}{2012}).

\bibitem[{\citenamefont{Yi et~al.}(2011)\citenamefont{Yi, Lu, Chu, Analytis,
  Sorini, Kemper, Moritz, Mo, Moore, Hashimoto et~al.}}]{Yi26042011}
\bibinfo{author}{\bibfnamefont{M.}~\bibnamefont{Yi}},
  \bibinfo{author}{\bibfnamefont{D.}~\bibnamefont{Lu}},
  \bibinfo{author}{\bibfnamefont{J.-H.} \bibnamefont{Chu}},
  \bibinfo{author}{\bibfnamefont{J.~G.} \bibnamefont{Analytis}},
  \bibinfo{author}{\bibfnamefont{A.~P.} \bibnamefont{Sorini}},
  \bibinfo{author}{\bibfnamefont{A.~F.} \bibnamefont{Kemper}},
  \bibinfo{author}{\bibfnamefont{B.}~\bibnamefont{Moritz}},
  \bibinfo{author}{\bibfnamefont{S.-K.} \bibnamefont{Mo}},
  \bibinfo{author}{\bibfnamefont{R.~G.} \bibnamefont{Moore}},
  \bibinfo{author}{\bibfnamefont{M.}~\bibnamefont{Hashimoto}},
  \bibnamefont{et~al.}, \bibinfo{journal}{Proceedings of the National Academy
  of Sciences} \textbf{\bibinfo{volume}{108}}, \bibinfo{pages}{6878}
  (\bibinfo{year}{2011}).

\bibitem[{\citenamefont{Kasahara et~al.}(2012)\citenamefont{Kasahara, Shi,
  Hashimoto, Tonegawa, Mizukami, Shibauchi, Sugimoto, Fukuda, Terashima,
  Nevidomskyy et~al.}}]{Kasahara}
\bibinfo{author}{\bibfnamefont{S.}~\bibnamefont{Kasahara}},
  \bibinfo{author}{\bibfnamefont{H.~J.} \bibnamefont{Shi}},
  \bibinfo{author}{\bibfnamefont{K.}~\bibnamefont{Hashimoto}},
  \bibinfo{author}{\bibfnamefont{S.}~\bibnamefont{Tonegawa}},
  \bibinfo{author}{\bibfnamefont{Y.}~\bibnamefont{Mizukami}},
  \bibinfo{author}{\bibfnamefont{T.}~\bibnamefont{Shibauchi}},
  \bibinfo{author}{\bibfnamefont{K.}~\bibnamefont{Sugimoto}},
  \bibinfo{author}{\bibfnamefont{T.}~\bibnamefont{Fukuda}},
  \bibinfo{author}{\bibfnamefont{T.}~\bibnamefont{Terashima}},
  \bibinfo{author}{\bibfnamefont{A.~H.} \bibnamefont{Nevidomskyy}},
  \bibnamefont{et~al.}, \bibinfo{journal}{Nature}
  \textbf{\bibinfo{volume}{486}}, \bibinfo{pages}{382} (\bibinfo{year}{2012}).

\bibitem[{\citenamefont{Fradkin et~al.}(2010)\citenamefont{Fradkin, Kivelson,
  Lawler, Eisenstein, and Mackenzie}}]{kivelson}
\bibinfo{author}{\bibfnamefont{E.}~\bibnamefont{Fradkin}},
  \bibinfo{author}{\bibfnamefont{S.~A.} \bibnamefont{Kivelson}},
  \bibinfo{author}{\bibfnamefont{M.~J.} \bibnamefont{Lawler}},
  \bibinfo{author}{\bibfnamefont{J.~P.} \bibnamefont{Eisenstein}},
  \bibnamefont{and} \bibinfo{author}{\bibfnamefont{A.~P.}
  \bibnamefont{Mackenzie}}, \bibinfo{journal}{Annual Review of Condensed Matter
  Physics} \textbf{\bibinfo{volume}{1}}, \bibinfo{pages}{153}
  (\bibinfo{year}{2010}).

\bibitem[{\citenamefont{de~la Cruz et~al.}(2008)\citenamefont{de~la Cruz,
  Huang, Lynn, Li, II, Zarestky, Mook, Chen, Luo, Wang et~al.}}]{delacruz}
\bibinfo{author}{\bibfnamefont{C.}~\bibnamefont{de~la Cruz}},
  \bibinfo{author}{\bibfnamefont{Q.}~\bibnamefont{Huang}},
  \bibinfo{author}{\bibfnamefont{J.~W.} \bibnamefont{Lynn}},
  \bibinfo{author}{\bibfnamefont{J.}~\bibnamefont{Li}},
  \bibinfo{author}{\bibfnamefont{W.~R.} \bibnamefont{II}},
  \bibinfo{author}{\bibfnamefont{J.~L.} \bibnamefont{Zarestky}},
  \bibinfo{author}{\bibfnamefont{H.~A.} \bibnamefont{Mook}},
  \bibinfo{author}{\bibfnamefont{G.~F.} \bibnamefont{Chen}},
  \bibinfo{author}{\bibfnamefont{J.~L.} \bibnamefont{Luo}},
  \bibinfo{author}{\bibfnamefont{N.~L.} \bibnamefont{Wang}},
  \bibnamefont{et~al.}, \bibinfo{journal}{Nature}
  \textbf{\bibinfo{volume}{453}}, \bibinfo{pages}{899} (\bibinfo{year}{2008}).

\bibitem[{\citenamefont{Huang et~al.}(2008)\citenamefont{Huang, Qiu, Bao,
  Green, Lynn, Gasparovic, Wu, Wu, and Chen}}]{PhysRevLett.101.257003}
\bibinfo{author}{\bibfnamefont{Q.}~\bibnamefont{Huang}},
  \bibinfo{author}{\bibfnamefont{Y.}~\bibnamefont{Qiu}},
  \bibinfo{author}{\bibfnamefont{W.}~\bibnamefont{Bao}},
  \bibinfo{author}{\bibfnamefont{M.~A.} \bibnamefont{Green}},
  \bibinfo{author}{\bibfnamefont{J.~W.} \bibnamefont{Lynn}},
  \bibinfo{author}{\bibfnamefont{Y.~C.} \bibnamefont{Gasparovic}},
  \bibinfo{author}{\bibfnamefont{T.}~\bibnamefont{Wu}},
  \bibinfo{author}{\bibfnamefont{G.}~\bibnamefont{Wu}}, \bibnamefont{and}
  \bibinfo{author}{\bibfnamefont{X.~H.} \bibnamefont{Chen}},
  \bibinfo{journal}{Phys. Rev. Lett.} \textbf{\bibinfo{volume}{101}},
  \bibinfo{pages}{257003} (\bibinfo{year}{2008}).

\bibitem[{\citenamefont{Chu et~al.}(2010)\citenamefont{Chu, Analytis, De~Greve,
  McMahon, Islam, Yamamoto, and Fisher}}]{Chu13082010}
\bibinfo{author}{\bibfnamefont{J.-H.} \bibnamefont{Chu}},
  \bibinfo{author}{\bibfnamefont{J.~G.} \bibnamefont{Analytis}},
  \bibinfo{author}{\bibfnamefont{K.}~\bibnamefont{De~Greve}},
  \bibinfo{author}{\bibfnamefont{P.~L.} \bibnamefont{McMahon}},
  \bibinfo{author}{\bibfnamefont{Z.}~\bibnamefont{Islam}},
  \bibinfo{author}{\bibfnamefont{Y.}~\bibnamefont{Yamamoto}}, \bibnamefont{and}
  \bibinfo{author}{\bibfnamefont{I.~R.} \bibnamefont{Fisher}},
  \bibinfo{journal}{Science} \textbf{\bibinfo{volume}{329}},
  \bibinfo{pages}{824} (\bibinfo{year}{2010}).

\bibitem[{\citenamefont{Dhital et~al.}(2012)\citenamefont{Dhital, Yamani, Tian,
  Zeretsky, Sefat, Wang, Birgeneau, and Wilson}}]{PhysRevLett.108.087001}
\bibinfo{author}{\bibfnamefont{C.}~\bibnamefont{Dhital}},
  \bibinfo{author}{\bibfnamefont{Z.}~\bibnamefont{Yamani}},
  \bibinfo{author}{\bibfnamefont{W.}~\bibnamefont{Tian}},
  \bibinfo{author}{\bibfnamefont{J.}~\bibnamefont{Zeretsky}},
  \bibinfo{author}{\bibfnamefont{A.~S.} \bibnamefont{Sefat}},
  \bibinfo{author}{\bibfnamefont{Z.}~\bibnamefont{Wang}},
  \bibinfo{author}{\bibfnamefont{R.~J.} \bibnamefont{Birgeneau}},
  \bibnamefont{and} \bibinfo{author}{\bibfnamefont{S.~D.}
  \bibnamefont{Wilson}}, \bibinfo{journal}{Phys. Rev. Lett.}
  \textbf{\bibinfo{volume}{108}}, \bibinfo{pages}{087001}
  (\bibinfo{year}{2012}).

\bibitem[{\citenamefont{Blomberg et~al.}(2012)\citenamefont{Blomberg, Kreyssig,
  Tanatar, Fernandes, Kim, Thaler, Schmalian, Bud'ko, Canfield, Goldman
  et~al.}}]{PhysRevB.85.144509}
\bibinfo{author}{\bibfnamefont{E.~C.} \bibnamefont{Blomberg}},
  \bibinfo{author}{\bibfnamefont{A.}~\bibnamefont{Kreyssig}},
  \bibinfo{author}{\bibfnamefont{M.~A.} \bibnamefont{Tanatar}},
  \bibinfo{author}{\bibfnamefont{R.~M.} \bibnamefont{Fernandes}},
  \bibinfo{author}{\bibfnamefont{M.~G.} \bibnamefont{Kim}},
  \bibinfo{author}{\bibfnamefont{A.}~\bibnamefont{Thaler}},
  \bibinfo{author}{\bibfnamefont{J.}~\bibnamefont{Schmalian}},
  \bibinfo{author}{\bibfnamefont{S.~L.} \bibnamefont{Bud'ko}},
  \bibinfo{author}{\bibfnamefont{P.~C.} \bibnamefont{Canfield}},
  \bibinfo{author}{\bibfnamefont{A.~I.} \bibnamefont{Goldman}},
  \bibnamefont{et~al.}, \bibinfo{journal}{Phys. Rev. B}
  \textbf{\bibinfo{volume}{85}}, \bibinfo{pages}{144509}
  (\bibinfo{year}{2012}).

\bibitem[{\citenamefont{Tanatar et~al.}(2009)\citenamefont{Tanatar, Kreyssig,
  Nandi, Ni, Bud'ko, Canfield, Goldman, and Prozorov}}]{PhysRevB.79.180508}
\bibinfo{author}{\bibfnamefont{M.~A.} \bibnamefont{Tanatar}},
  \bibinfo{author}{\bibfnamefont{A.}~\bibnamefont{Kreyssig}},
  \bibinfo{author}{\bibfnamefont{S.}~\bibnamefont{Nandi}},
  \bibinfo{author}{\bibfnamefont{N.}~\bibnamefont{Ni}},
  \bibinfo{author}{\bibfnamefont{S.~L.} \bibnamefont{Bud'ko}},
  \bibinfo{author}{\bibfnamefont{P.~C.} \bibnamefont{Canfield}},
  \bibinfo{author}{\bibfnamefont{A.~I.} \bibnamefont{Goldman}},
  \bibnamefont{and} \bibinfo{author}{\bibfnamefont{R.}~\bibnamefont{Prozorov}},
  \bibinfo{journal}{Phys. Rev. B} \textbf{\bibinfo{volume}{79}},
  \bibinfo{pages}{180508} (\bibinfo{year}{2009}),
  \urlprefix\url{http://link.aps.org/doi/10.1103/PhysRevB.79.180508}.

\bibitem[{\citenamefont{Dhital et~al.}(2014)\citenamefont{Dhital, Hogan,
  Yamani, Birgeneau, Tian, Matsuda, Sefat, Wang, and
  Wilson}}]{PhysRevB.89.214404}
\bibinfo{author}{\bibfnamefont{C.}~\bibnamefont{Dhital}},
  \bibinfo{author}{\bibfnamefont{T.}~\bibnamefont{Hogan}},
  \bibinfo{author}{\bibfnamefont{Z.}~\bibnamefont{Yamani}},
  \bibinfo{author}{\bibfnamefont{R.~J.} \bibnamefont{Birgeneau}},
  \bibinfo{author}{\bibfnamefont{W.}~\bibnamefont{Tian}},
  \bibinfo{author}{\bibfnamefont{M.}~\bibnamefont{Matsuda}},
  \bibinfo{author}{\bibfnamefont{A.~S.} \bibnamefont{Sefat}},
  \bibinfo{author}{\bibfnamefont{Z.}~\bibnamefont{Wang}}, \bibnamefont{and}
  \bibinfo{author}{\bibfnamefont{S.~D.} \bibnamefont{Wilson}},
  \bibinfo{journal}{Phys. Rev. B} \textbf{\bibinfo{volume}{89}},
  \bibinfo{pages}{214404} (\bibinfo{year}{2014}).

\bibitem[{\citenamefont{Ren et~al.}(2015)\citenamefont{Ren, Duan, Hu, Li,
  Zhang, Luo, Dai, and Li}}]{PhysRevLett.115.197002}
\bibinfo{author}{\bibfnamefont{X.}~\bibnamefont{Ren}},
  \bibinfo{author}{\bibfnamefont{L.}~\bibnamefont{Duan}},
  \bibinfo{author}{\bibfnamefont{Y.}~\bibnamefont{Hu}},
  \bibinfo{author}{\bibfnamefont{J.}~\bibnamefont{Li}},
  \bibinfo{author}{\bibfnamefont{R.}~\bibnamefont{Zhang}},
  \bibinfo{author}{\bibfnamefont{H.}~\bibnamefont{Luo}},
  \bibinfo{author}{\bibfnamefont{P.}~\bibnamefont{Dai}}, \bibnamefont{and}
  \bibinfo{author}{\bibfnamefont{Y.}~\bibnamefont{Li}}, \bibinfo{journal}{Phys.
  Rev. Lett.} \textbf{\bibinfo{volume}{115}}, \bibinfo{pages}{197002}
  (\bibinfo{year}{2015}).

\bibitem[{\citenamefont{Tomi\ifmmode~\acute{c}\else \'{c}\fi{}
  et~al.}(2013)\citenamefont{Tomi\ifmmode~\acute{c}\else \'{c}\fi{}, Jeschke,
  Fernandes, and Valent\'{\i}}}]{PhysRevB.87.174503}
\bibinfo{author}{\bibfnamefont{M.}~\bibnamefont{Tomi\ifmmode~\acute{c}\else
  \'{c}\fi{}}}, \bibinfo{author}{\bibfnamefont{H.~O.} \bibnamefont{Jeschke}},
  \bibinfo{author}{\bibfnamefont{R.~M.} \bibnamefont{Fernandes}},
  \bibnamefont{and}
  \bibinfo{author}{\bibfnamefont{R.}~\bibnamefont{Valent\'{\i}}},
  \bibinfo{journal}{Phys. Rev. B} \textbf{\bibinfo{volume}{87}},
  \bibinfo{pages}{174503} (\bibinfo{year}{2013}).

\bibitem[{\citenamefont{Tanatar et~al.}(2010)\citenamefont{Tanatar, Blomberg,
  Kreyssig, Kim, Ni, Thaler, Bud'ko, Canfield, Goldman, Mazin
  et~al.}}]{PhysRevB.81.184508}
\bibinfo{author}{\bibfnamefont{M.~A.} \bibnamefont{Tanatar}},
  \bibinfo{author}{\bibfnamefont{E.~C.} \bibnamefont{Blomberg}},
  \bibinfo{author}{\bibfnamefont{A.}~\bibnamefont{Kreyssig}},
  \bibinfo{author}{\bibfnamefont{M.~G.} \bibnamefont{Kim}},
  \bibinfo{author}{\bibfnamefont{N.}~\bibnamefont{Ni}},
  \bibinfo{author}{\bibfnamefont{A.}~\bibnamefont{Thaler}},
  \bibinfo{author}{\bibfnamefont{S.~L.} \bibnamefont{Bud'ko}},
  \bibinfo{author}{\bibfnamefont{P.~C.} \bibnamefont{Canfield}},
  \bibinfo{author}{\bibfnamefont{A.~I.} \bibnamefont{Goldman}},
  \bibinfo{author}{\bibfnamefont{I.~I.} \bibnamefont{Mazin}},
  \bibnamefont{et~al.}, \bibinfo{journal}{Phys. Rev. B}
  \textbf{\bibinfo{volume}{81}}, \bibinfo{pages}{184508}
  (\bibinfo{year}{2010}).

\bibitem[{\citenamefont{Lu et~al.}(2014)\citenamefont{Lu, Park, Zhang, Luo,
  Nevidomskyy, Si, and Dai}}]{Lu08082014}
\bibinfo{author}{\bibfnamefont{X.}~\bibnamefont{Lu}},
  \bibinfo{author}{\bibfnamefont{J.~T.} \bibnamefont{Park}},
  \bibinfo{author}{\bibfnamefont{R.}~\bibnamefont{Zhang}},
  \bibinfo{author}{\bibfnamefont{H.}~\bibnamefont{Luo}},
  \bibinfo{author}{\bibfnamefont{A.~H.} \bibnamefont{Nevidomskyy}},
  \bibinfo{author}{\bibfnamefont{Q.}~\bibnamefont{Si}}, \bibnamefont{and}
  \bibinfo{author}{\bibfnamefont{P.}~\bibnamefont{Dai}},
  \bibinfo{journal}{Science} \textbf{\bibinfo{volume}{345}},
  \bibinfo{pages}{657} (\bibinfo{year}{2014}).

\bibitem[{\citenamefont{Rotundu et~al.}(2010)\citenamefont{Rotundu, Freelon,
  Forrest, Wilson, Valdivia, Pinuellas, Kim, Kim, Islam, Bourret-Courchesne
  et~al.}}]{PhysRevB.82.144525}
\bibinfo{author}{\bibfnamefont{C.~R.} \bibnamefont{Rotundu}},
  \bibinfo{author}{\bibfnamefont{B.}~\bibnamefont{Freelon}},
  \bibinfo{author}{\bibfnamefont{T.~R.} \bibnamefont{Forrest}},
  \bibinfo{author}{\bibfnamefont{S.~D.} \bibnamefont{Wilson}},
  \bibinfo{author}{\bibfnamefont{P.~N.} \bibnamefont{Valdivia}},
  \bibinfo{author}{\bibfnamefont{G.}~\bibnamefont{Pinuellas}},
  \bibinfo{author}{\bibfnamefont{A.}~\bibnamefont{Kim}},
  \bibinfo{author}{\bibfnamefont{J.-W.} \bibnamefont{Kim}},
  \bibinfo{author}{\bibfnamefont{Z.}~\bibnamefont{Islam}},
  \bibinfo{author}{\bibfnamefont{E.}~\bibnamefont{Bourret-Courchesne}},
  \bibnamefont{et~al.}, \bibinfo{journal}{Phys. Rev. B}
  \textbf{\bibinfo{volume}{82}}, \bibinfo{pages}{144525}
  (\bibinfo{year}{2010}).

\bibitem[{\citenamefont{Song et~al.}(2013)\citenamefont{Song, Carr, Lu, Zhang,
  Sims, Luttrell, Chi, Zhao, Lynn, and Dai}}]{PhysRevB.87.184511}
\bibinfo{author}{\bibfnamefont{Y.}~\bibnamefont{Song}},
  \bibinfo{author}{\bibfnamefont{S.~V.} \bibnamefont{Carr}},
  \bibinfo{author}{\bibfnamefont{X.}~\bibnamefont{Lu}},
  \bibinfo{author}{\bibfnamefont{C.}~\bibnamefont{Zhang}},
  \bibinfo{author}{\bibfnamefont{Z.~C.} \bibnamefont{Sims}},
  \bibinfo{author}{\bibfnamefont{N.~F.} \bibnamefont{Luttrell}},
  \bibinfo{author}{\bibfnamefont{S.}~\bibnamefont{Chi}},
  \bibinfo{author}{\bibfnamefont{Y.}~\bibnamefont{Zhao}},
  \bibinfo{author}{\bibfnamefont{J.~W.} \bibnamefont{Lynn}}, \bibnamefont{and}
  \bibinfo{author}{\bibfnamefont{P.}~\bibnamefont{Dai}},
  \bibinfo{journal}{Phys. Rev. B} \textbf{\bibinfo{volume}{87}},
  \bibinfo{pages}{184511} (\bibinfo{year}{2013}).

\bibitem[{\citenamefont{Liang et~al.}(2011)\citenamefont{Liang, Nakajima,
  Kihou, Tomioka, Ito, Lee, Kito, Iyo, Eisaki, Kakeshita
  et~al.}}]{Liang2011418}
\bibinfo{author}{\bibfnamefont{T.}~\bibnamefont{Liang}},
  \bibinfo{author}{\bibfnamefont{M.}~\bibnamefont{Nakajima}},
  \bibinfo{author}{\bibfnamefont{K.}~\bibnamefont{Kihou}},
  \bibinfo{author}{\bibfnamefont{Y.}~\bibnamefont{Tomioka}},
  \bibinfo{author}{\bibfnamefont{T.}~\bibnamefont{Ito}},
  \bibinfo{author}{\bibfnamefont{C.}~\bibnamefont{Lee}},
  \bibinfo{author}{\bibfnamefont{H.}~\bibnamefont{Kito}},
  \bibinfo{author}{\bibfnamefont{A.}~\bibnamefont{Iyo}},
  \bibinfo{author}{\bibfnamefont{H.}~\bibnamefont{Eisaki}},
  \bibinfo{author}{\bibfnamefont{T.}~\bibnamefont{Kakeshita}},
  \bibnamefont{et~al.}, \bibinfo{journal}{Journal of Physics and Chemistry of
  Solids} \textbf{\bibinfo{volume}{72}}, \bibinfo{pages}{418 }
  (\bibinfo{year}{2011}), ISSN \bibinfo{issn}{0022-3697}.

\bibitem[{\citenamefont{{Lu} et~al.}(2015)\citenamefont{{Lu}, {Keller},
  {Zhang}, {Song}, {Park}, {Luo}, {Li}, and {Dai}}}]{2015arXiv150704191L}
\bibinfo{author}{\bibfnamefont{X.}~\bibnamefont{{Lu}}},
  \bibinfo{author}{\bibfnamefont{T.}~\bibnamefont{{Keller}}},
  \bibinfo{author}{\bibfnamefont{W.}~\bibnamefont{{Zhang}}},
  \bibinfo{author}{\bibfnamefont{Y.}~\bibnamefont{{Song}}},
  \bibinfo{author}{\bibfnamefont{J.~T.} \bibnamefont{{Park}}},
  \bibinfo{author}{\bibfnamefont{H.}~\bibnamefont{{Luo}}},
  \bibinfo{author}{\bibfnamefont{S.}~\bibnamefont{{Li}}}, \bibnamefont{and}
  \bibinfo{author}{\bibfnamefont{P.}~\bibnamefont{{Dai}}},
  \bibinfo{journal}{ArXiv e-prints}  (\bibinfo{year}{2015}),
  \eprint{1507.04191}.

\bibitem[{Sup()}]{Supplemental}
\bibinfo{note}{See Supplemental Information for additional details.}

\bibitem[{\citenamefont{Fernandes et~al.}(2010)\citenamefont{Fernandes,
  VanBebber, Bhattacharya, Chandra, Keppens, Mandrus, McGuire, Sales, Sefat,
  and Schmalian}}]{PhysRevLett.105.157003}
\bibinfo{author}{\bibfnamefont{R.~M.} \bibnamefont{Fernandes}},
  \bibinfo{author}{\bibfnamefont{L.~H.} \bibnamefont{VanBebber}},
  \bibinfo{author}{\bibfnamefont{S.}~\bibnamefont{Bhattacharya}},
  \bibinfo{author}{\bibfnamefont{P.}~\bibnamefont{Chandra}},
  \bibinfo{author}{\bibfnamefont{V.}~\bibnamefont{Keppens}},
  \bibinfo{author}{\bibfnamefont{D.}~\bibnamefont{Mandrus}},
  \bibinfo{author}{\bibfnamefont{M.~A.} \bibnamefont{McGuire}},
  \bibinfo{author}{\bibfnamefont{B.~C.} \bibnamefont{Sales}},
  \bibinfo{author}{\bibfnamefont{A.~S.} \bibnamefont{Sefat}}, \bibnamefont{and}
  \bibinfo{author}{\bibfnamefont{J.}~\bibnamefont{Schmalian}},
  \bibinfo{journal}{Phys. Rev. Lett.} \textbf{\bibinfo{volume}{105}},
  \bibinfo{pages}{157003} (\bibinfo{year}{2010}).

\bibitem[{\citenamefont{Inoue et~al.}(2012)\citenamefont{Inoue, Yamakawa, and
  Kontani}}]{PhysRevB.85.224506}
\bibinfo{author}{\bibfnamefont{Y.}~\bibnamefont{Inoue}},
  \bibinfo{author}{\bibfnamefont{Y.}~\bibnamefont{Yamakawa}}, \bibnamefont{and}
  \bibinfo{author}{\bibfnamefont{H.}~\bibnamefont{Kontani}},
  \bibinfo{journal}{Phys. Rev. B} \textbf{\bibinfo{volume}{85}},
  \bibinfo{pages}{224506} (\bibinfo{year}{2012}).

\bibitem[{\citenamefont{Khan et~al.}(2013)\citenamefont{Khan, Alam, and
  Johnson}}]{PhysRevB.88.184515}
\bibinfo{author}{\bibfnamefont{S.~N.} \bibnamefont{Khan}},
  \bibinfo{author}{\bibfnamefont{A.}~\bibnamefont{Alam}}, \bibnamefont{and}
  \bibinfo{author}{\bibfnamefont{D.~D.} \bibnamefont{Johnson}},
  \bibinfo{journal}{Phys. Rev. B} \textbf{\bibinfo{volume}{88}},
  \bibinfo{pages}{184515} (\bibinfo{year}{2013}).

\bibitem[{\citenamefont{Niedziela et~al.}(2012)\citenamefont{Niedziela,
  McGuire, and Egami}}]{PhysRevB.86.174113}
\bibinfo{author}{\bibfnamefont{J.~L.} \bibnamefont{Niedziela}},
  \bibinfo{author}{\bibfnamefont{M.~A.} \bibnamefont{McGuire}},
  \bibnamefont{and} \bibinfo{author}{\bibfnamefont{T.}~\bibnamefont{Egami}},
  \bibinfo{journal}{Phys. Rev. B} \textbf{\bibinfo{volume}{86}},
  \bibinfo{pages}{174113} (\bibinfo{year}{2012}).

\bibitem[{\citenamefont{Man et~al.}(2015)\citenamefont{Man, Lu, Chen, Zhang,
  Zhang, Luo, Kulda, Ivanov, Keller, Morosan et~al.}}]{Man}
\bibinfo{author}{\bibfnamefont{H.}~\bibnamefont{Man}},
  \bibinfo{author}{\bibfnamefont{X.}~\bibnamefont{Lu}},
  \bibinfo{author}{\bibfnamefont{J.~S.} \bibnamefont{Chen}},
  \bibinfo{author}{\bibfnamefont{R.}~\bibnamefont{Zhang}},
  \bibinfo{author}{\bibfnamefont{W.}~\bibnamefont{Zhang}},
  \bibinfo{author}{\bibfnamefont{H.}~\bibnamefont{Luo}},
  \bibinfo{author}{\bibfnamefont{J.}~\bibnamefont{Kulda}},
  \bibinfo{author}{\bibfnamefont{A.}~\bibnamefont{Ivanov}},
  \bibinfo{author}{\bibfnamefont{T.}~\bibnamefont{Keller}},
  \bibinfo{author}{\bibfnamefont{E.}~\bibnamefont{Morosan}},
  \bibnamefont{et~al.}, \bibinfo{journal}{Phys. Rev. B}
  \textbf{\bibinfo{volume}{92}}, \bibinfo{pages}{134521}
  (\bibinfo{year}{2015}),
  \urlprefix\url{http://link.aps.org/doi/10.1103/PhysRevB.92.134521}.

\end{thebibliography}
%
%
\end{document}